\documentclass{ws-procs9x6}
\usepackage{amssymb,amsfonts,amsmath}
\usepackage{epsfig,graphicx}\usepackage{mathrsfs}

\newcommand{\grp}[1]{\mathsf{#1}}

\newcommand{\spc}[1]{\mathscr{#1}}
\newcommand{\Span}{{\mathsf{Span}}}

\newcommand{\B}{{\mathcal B}} 
\def\qed{$\blacksquare$ \newline}
\def\map#1{\mathcal{#1}}
\def\Proof{{\bf Proof.~}}

\begin{document}
\title{Confusability graphs for symmetric sets of quantum states}

\author{Giulio Chiribella} 
\address{Center for Quantum Information, Institute for Interdisciplinary Information Sciences, Tsinghua University, Beijing, 100084, China\\
Email: gchiribella@mail.tinsghua.edu.cn
} 
\author{Yuxiang Yang}
\address{Department of Physics, Tsinghua University, Beijing, 100084, China\\
Email: yangyx09@mails.tsinghua.edu.cn}

\begin{abstract}
For a set of quantum states generated by the action of a group, we consider the graph obtained by considering two group elements adjacent whenever the corresponding states are non-orthogonal.  We analyze the structure of the connected components of the graph and show two  applications to the optimal estimation of an unknown group action and to the search for decoherence free subspaces of quantum channels with symmetry.     
\end{abstract}

\section{Introduction}
A fundamental feature of quantum theory is the impossibility to reliably distinguish between non-orthogonal pure states  \cite{helstrom}.  This impossibility is at the root of the no-cloning theorem \cite{dieks82,woot82} and of the information-disturbance trade-offs that guarantee the security of quantum cryptography \cite{gisin02}.   Non-orthogonality can be used to partition a set of pure states into subsets exhibiting gnuine quantum features.  Consider a map  $\varphi$ 
 that encodes elements of a set  $\mathsf X$  into unit vectors a Hilbert space $\spc H$ 
 \begin{align}
 \varphi   :  \mathsf X  \to \spc H \qquad  x \in  \mathsf X  \mapsto  |\varphi_x\rangle
  \in  \spc H 
 \end{align} 
The map $\varphi$ endows $\mathsf X$ with a graph structure: 
\begin{definition}   
The \emph{confusability graph} associated to $\varphi$, denoted by  $ \Gamma ({\grp X}, \spc H, \varphi)$,  is the  graph    where $\grp X $ is the set of vertices  and two vertices  $x,y\in \grp X $ are connected by an edge whenever   $\langle
  \varphi_x  |  \varphi_y\rangle
  \not = 0$.    
\end{definition}
By this definition, two vertices $x,y\in \grp X $ are adjacent if and only if the corresponding states $|\varphi_x\rangle
,  |\varphi_y\rangle
 \in \spc H$ are confusable  (i.e. non-orthogonal). A vertex $x$ is connected to a vertex $y$ if and only if we can go from the state $|\varphi_x\rangle
$ to the state $|\varphi_y\rangle
$ through a path of confusable states. 

Denoting by $\Gamma_k$, $k\in  \grp K$  the connected components of $\Gamma (  \grp X, \spc H, \varphi)$ and by $\grp X_k$ the subset of vertices belonging to $\Gamma_k$, we  can partition the set of states $\varphi (\grp X) =  \{  |\varphi_x\rangle
 ~|~  x\in \mathsf X\}$ into mutually orthogonal subsets   $\varphi (\grp X_k)  = \{  |\varphi_x\rangle
 ~|~  x\in \mathsf X_k\}$.     
Intuitively, the subsets  $\varphi (\grp X_k)$ identify sectors of the Hilbert space $\spc H$ where the encoding $\varphi$ exhibits genuinely quantum features.  More precisely, one can define the mutually orthogonal subspaces  
\begin{align}\label{subspaces}
S_k  :  =  \Span  \{   |\varphi_x\rangle
 ~|~  x\in \mathsf X_k   \} \qquad   k \in  \grp K 
\end{align}  
and the corresponding orthogonal projectors $P_k,   k\in\mathsf K$.   With this definition, the linear map  $\map D$  on trace-class operators   defined by 
\begin{align}\label{deco}
\map D (\rho)  =  \sum_{k\in \grp K}   P_k  \rho P_k
\end{align} 
 does not disturb states in the set $\varphi(\grp X)$, that is, $\map D  ( |\varphi_x\rangle
\langle
  \varphi_x| )  =   |\varphi_x\rangle
\langle
  \varphi_x|$ for every $x\in  \mathsf X$.   The map $\map D$ represents a partial decoherence process that preserves the off-diagonal elements of a density matrix only within the subspaces $S_k$.  Equivalently, we can interpret $\map D$ as the result of a L\"uders measurement that extracts information about 
 the orthogonal subspaces $S_k$ without disturbing states that have support inside these subspaces. 
   
In this paper, we consider confusability graphs generated by the action of a group $\grp G$. The map $\varphi$ will be of the form 
\begin{align}
\varphi :  \grp G  \to \spc H   \qquad   g \in  \grp G   \mapsto  |\varphi_g\rangle
  :  =   U_g  |\varphi\rangle
   ,    
\end{align}  
where $|\varphi\rangle
  \in  \spc H$  is a unit vector and $U:  \grp G  \to \B (\spc H),  g  \mapsto U_g $  is a unitary projective representation of $\grp G$.   In this special case, the confusability graph will be denoted by $\Gamma(  \grp G, \spc H,  U, |\varphi\rangle
   )$, to stress that the map $\varphi$ is now specified by the choice of a representation $U$ and of an input state $|\varphi\rangle
$.   In section {\ref{sec:group}}, we will show that the component of the graph connected to the identity is a subgroup of $\grp G$ and that the other components are left-cosets.   We then show the implications of this structure for the problem of quantum estimation of an unknown group action  (section \ref{sec:est}) and for the problem of finding decoherence free subspaces of quantum channels with symmetry (section \ref{sec:decofree}).  In particular, we will show that a covariant channel that preserves a single pure state $|\varphi\rangle
  \in\spc H$  will automatically preserve any state with support in one of the subspaces $S_k$. In other words, all the subspaces $S_k$ will be automatically \emph{decoherence free}  \cite{palma96,duan97,zan97,lid98,choi06}.

Before presenting our results, we would like to mention their connection to previous related works.     
An early appearance of a confusability graph is the \emph{orthonormal representation of a graph} introduced by  Lov\'{a}sz \cite{lov79}.  This notion, which is central in the study of zero error communication(see e.g. Refs.\cite{beigi10,cubitt10}), was recently generalized to the quantum case by Duan, Severini, and Winter \cite{duan11}.  In zero-error communication, the problem is to find maximal sets of disconnected points in the graph (by definition, the messages that are communicated must not be confusable).  Our discussion will take a complementary point of view, focussing instead on the connected components and on their role in quantum estimation and error correction.  

\section{Group theoretic structure of the confusability graph}\label{sec:group}
Consider a confusability graph  $ \Gamma   ( \grp G, \spc H,  U,  |\psi\rangle
)$.     By definition, two group elements  $g\in \grp G$ and $h\in\grp G$ are connected, denoted by $\sim$, if and only if there exists a finite path $\{   g_i  \in  \grp G\}_{i=1}^N $    such that $g_1  =  g$,  $g_N  =  h$,  and $  \prod_{i=1}^{N-1}   \langle
  \varphi_{g_{i}}   |   \varphi_{g_{i+1}}\rangle
   \not  = 0 $.   It is then easy to prove the following

\begin{proposition}\label{prop:group}  Let  $ \Gamma   ( \grp G, \spc H,  U,  |\varphi\rangle
)$  be a confusability graph. Then,
\begin{enumerate}
\item the connected component  $\grp H :  =  \{   h \in  \grp G  ~|~   h \sim e\}$   is a subgroup of $\grp G$ 
\item the connected components of $ \Gamma   ( \grp G, \spc H,  U,  |\varphi\rangle
)$ are the left cosets of $\grp G$  with respect to $\grp H$.  
\end{enumerate}     
\end{proposition}

\Proof If $g \sim g'$, then  $h g  \sim hg'$ for every $h\in\grp G$.   Using this fact it is immediate to see that $\grp H$ is a group.  It is also clear that $g \sim g'$ if and only if $g^{-1}  g' \sim e$, that is, if and only if  $g^{-1}  g' $ belongs to $\grp H$.  Hence, the connected component $\grp H_{g}  :  \{ g'  \in  \grp G~|~  g'  \sim g \}$ is the left coset  $\grp H_g =  g \grp H$. 
   \qed 
Note that for a continuous representation of a topological group, the subgroup $\grp H$ always contains the component of the group topologically connected to the identity.  However, the notion of connectedness that we use here makes sense also for finite groups and in cases where no topology is given a priori.

An important family of input states, containing the best possible input states for the estimation of an unknown group element $g\in  \grp G$, is the family of the \emph{class states} \cite{proc11}:  a class state $|\varphi\rangle
  \in  \spc H $ can be concisely defined as a vector such that the characteristic function\cite{marv11} $  f(g)  :=  \langle
 \varphi  |   U_g  |\varphi\rangle
 $ is a class function, namely a function such that $f(h^{-1}  g  h  )  =  f(g) $ for every $g,h \in \grp G$.  For this particular type of states we have the condition

\begin{proposition}\label{prop:norm}  Let  $ \Gamma   ( \grp G, \spc H,  U,  |\varphi\rangle
)$  be the confusability graph of a class state $|\varphi\rangle
  \in  \spc H$. Then, the connected component  $\grp H$   is a  \emph{normal} subgroup of $\grp G$. 
\end{proposition}

\Proof For a class state,  $g \sim g'$ implies   $g  h  \sim g' h$ for every $h\in\grp G$.  By the same proof of proposition  \ref{prop:group}, we then obtain that the connected components are right cosets of $\grp G$.   Hence, we must have   $g \grp H = \grp H  g$ for every $g\in\grp G$, because    $g \grp H $ and  $\grp H  g$ are two intersecting connected components.   \qed  

In the case of class states, the  decoherence map in Eq. (\ref{deco}) can be interpreted as the result of an projective measurement with outcome in the quotient group $\grp G/\grp H$.  

\section{Application to  quantum estimation}\label{sec:est}

The group-theoretic structure of the confusability graph has an immediate application to the problem of group parameter estimation   \cite{helstrom,holevo,chiribella05,proc11,giov06,buzek99,hayashi12}.    Suppose that a quantum system with Hilbert space $\spc H$,  initialized in a pure state $|\psi\rangle
  \in \spc H$, undergoes an unknown unitary transformation  $U_g$.   The problem is to find the best quantum measurement, represented by \emph{positive operator valued measure  (POVM)}   $P(d \hat g)$ in order to estimate $g$ with maximum precision.      The figure of merit is the minimization of a cost functional
\begin{equation}\label{cost}
c(\rho, P):=   \sup_{g\in\grp G}   \left\{     \int_{\grp G} c(\hat{g},g)  ~  \langle
 \varphi_g  |   P(d\hat{g}  |\varphi_g\rangle
    \right\},
\end{equation}
where $c(\hat{g},g)$ is a cost function satisfying the property  $c(h\hat{g},hg)  = c(\hat{g},g), \forall \hat g, g ,h\in\grp G$.   In this problem, it is well known that the optimization can be restricted to the set of \emph{covariant POVMs}  \cite{helstrom,holevo}, making the problem significantly simpler.     Inspecting the structure of the confusability graph, we can further simplify the problem, reducing the search of the optimal POVM to the search of an optimal  POVM for the lower-dimensional subspace $S_{\grp H}$, the subspace in Eq. (\ref{subspaces}) generated by the group elements in  $\grp H$.  Indeed,  we can write the unknown group element $g$ as $g  =   \tilde g h$, where $h$ belongs to $\grp H$ and $\tilde g$ is a fixed representative of the coset $g\grp H$.  To estimate $g$,   we can first identify $\tilde g$ by performing a projective measurement on the subspaces $  S_k$,  $k\in\grp G/\grp H$ defined in Eq. (\ref{subspaces}) (this measurement does not disturb the quantum state $|\varphi_g\rangle
$).   Then we can apply the unitary $U_{\tilde g}^\dag$ to the state $ |\varphi_g\rangle
$, turning it into the state $|\varphi_h\rangle
\in S_{\grp H}$.   Besides the reduction of the dimensionality, this procedure has the additional bonus that the estimation problem inside $S_{\grp H}$ still has the form   of Eq. (\ref{cost}), with the group $\grp G$ now replaced by its subgroup $\grp H$.   Hence, the problem of finding the optimal $\grp G$-covariant POVM on $\spc H$ is reduced to the problem of finding the optimal $\grp H$-covariant POVM on  $ S_{\grp H}$.

Note that the trick of identifying the group element $\tilde g$ by a projective measurement on the orthogonal subspaces $\{S_k\}_{k\in\grp G/\grp H}$ is reminiscent of the syndrome measurement in the stabilizer formalism \cite{gott96,gott97}.   This is not by chance, as one can rephrase several features of quantum  error correction in terms of confusability graphs associated to subspaces.

%\footnote{In the case of a subspace $\spc H_0  \subset \spc H$ one can define the confusability graph  $\Gamma  (\grp G,\spc H , U, \spc H_0)$ as the graph whose set of vertices is $\grp G$ and whose set of edges is the union of the union of the sets of vertices of the graphs $\Gamma  (\grp G,\spc H , U, \spc H_0)$.  In this case, the orthogonal subspace $S_k$ associated to the coset $  k\in  \grp  G/\grp H $   can be defined as $S_k  :=  \Span\{  U_g  |\varphi\rangle
 % ~|~  |\varphi \rangle
  %\in \spc H_0,  g\in   k  \}$. }       

\section{ Decoherence free subspaces of covariant channels }\label{sec:decofree}
The confusability graph has a remarkable application in the search for decoherence free subspaces of quantum channels with group symmetry.    
Consider a quantum channel $\map C$, namely a completely positive trace-preserving normal map transforming trace-class operators on $\spc H$.     The  channel $\map C$ is said to be \emph{covariant} if it satisfies the relation $\map C \circ \map U_g  = \map U_g \circ\map C, \forall  g\in\grp G$, where $\map U_g$ is the unitary channel defined by $\map U_g (\rho)  :=  U_g \rho U_g^\dag$.   
The following proposition can be used to simplify the search for decoherence free subspaces of covariant channels: 
    \begin{proposition}\label{prop:covchan} {\bf (Decoherence free subspaces of covariant channels)}
Suppose that the covariant channel  $\map C$ has a pure fixed point, i.e. that there is a pure state  $| \psi \rangle
  \in  \spc H$ such that $\map C (|\psi\rangle
\langle
\psi|)   = |\psi\rangle
\langle
\psi| $. Then, each subspace  $S_k$, corresponding to a connected component of the confusability graph $   \Gamma  (  \grp G,  \spc H,  U,  |\varphi\rangle
)$ is decoherence free.          
\end{proposition}

The proof is based on the observation that, due to covariance,  one has $\map C(|\psi_g\rangle
\langle
\psi_g|)=|\psi_g\rangle
\langle
\psi_g|$ for every $g \in \grp G$.     The  thesis follows by combining this fact with a simple lemma, which holds for arbitrary confusability graphs:   

\begin{lemma}\label{lem:dfs}
Suppose that the  projectors on the pure states $\{ \varphi_x  ~|~  x\in\grp X \} $  are fixed points of the  channel  $\map C$.  Then, each subspace $S_k$  associated to a connected component of the confusability  graph $\Gamma (\grp X, \spc H, \varphi)$  is decoherence free. 
\end{lemma}

The lemma can be proved as a corollary of a more general result about the algebraic structure of the fixed points of quantum channels \cite{koh08,koh10}.   However, the lemma has also a very elementary proof, similar to the proof of the no-cloning theorem \cite{dieks82,woot82}:  

{\bf Proof of lemma \ref{lem:dfs}}    Consider a unitary dilation of the channel $\map C$, given by $\map C(\rho)=\operatorname{Tr}
_{\spc H_E}[U  ( \rho\otimes|\eta\rangle
\langle
\eta| ) U^{\dagger}]$ for some Hilbert space $\spc H_E$, some unitary operator  $U \in  \B (\spc H_E)$, and some unit vector $|\eta\rangle
  \in  \spc H_E$.    Since the pure state $|\varphi_x\rangle
$  is a fixed point of $\map C$   we must have 
$ U|\varphi_x\rangle
|\eta\rangle
=|\varphi_x\rangle
|\eta_x\rangle
$, where $|\eta_x\rangle
\in\spc H_E$ is a unit vector possibly depending on $x$.  For two points $x,y\in  \grp X$, we then have    $\langle
 \varphi_x  |  \varphi_y\rangle
  =  \langle
 \varphi_x  |  \varphi_y\rangle
  \langle
 \eta_x|  \eta_y\rangle
$.      
  Now, if $\langle
 \varphi_x |\varphi_y \rangle
  \not = 0 $, we must have $|\eta_x\rangle
  =  |\eta_y\rangle
$.  Hence, the vector $|\eta_x\rangle
$ is constant on the connected components of the confusability graph $\Gamma ( \grp X, \spc H, \varphi )$:  $|\eta_x\rangle
  =  |\eta_{x'}  \equiv   |\eta_k\rangle
$ for every $x,x'\in \grp X_k $.   Taking a generic linear combination $|\psi\rangle
   =  \sum_{x  \in \grp X_k}    c_i  |\varphi_x\rangle
  \in  S_k$ we have  $U|\psi\rangle
|\eta\rangle
=|\psi\rangle
|\eta_k\rangle
$, and, therefore $\map C (  |\psi\rangle
\langle
  \psi|)  =  |\psi\rangle
\langle
\psi|$.   
\qed
  
In addition to simplifying the search for decoherence free subspaces, proposition \ref{prop:covchan} places a constraint on the covariant channels that can achieve a desired task---such as quantum teleportation \cite{bent93}---that involves the preservation of a pure state.  For example, a covariant channel that teleports a single state $|\varphi\rangle
$  must be able teleport all the states in the subspaces $S_k$ corresponding to connected components of the confusability graph $\Gamma(\grp G , \spc H, U,|\varphi\rangle
)$. 
The connected component $S_\grp H$ can easily be the whole Hilbert space:  for a teleportation protocol that is covariant under time-translations \cite{clocktele}, assuming a non-degenerate Hamiltonian, the ability to teleport a quantum state  $|\varphi\rangle
  = \sum_{n=0}^{\infty}  c_n  |n\rangle
 ,   ~  c_n  \not = 0,\forall n$ (and hence the whole trajectory   $|\varphi_t\rangle
  = \sum_{n=0}^{\infty}  c_n   e^{i  \omega_n  t} |n\rangle
 $, with $\omega_n  \not =  \omega_{n'}$ for $n \not  =  n'$) is equivalent to the ability to teleport an arbitrary quantum state $|\psi\rangle
\in  \spc H$.

\section{Conclusion}

In this short note we discussed the structure of the confusability graph for a set of pure states generated by  a group action, providing two applications of this structure to quantum estimation and to the search for decoherence free subspaces of quantum channels  with symmetry.  

%We expect that the study of the connected components of a distinguishability graph will have more applications, especially in relation to quantum error correction.     
 
 \medskip 
 \section*{Acknowledgement} 
This work is supported  the National Basic Research Program of China (973) 2011CBA00300 (2011CBA00301) and by the National Natural Science Foundation of China through Grants 61033001 and  61061130540.

\bibliographystyle{ws-procs9x6}
\bibliography{ws-pro-sample}

\end{document}